\documentclass[aps,prl,twocolumn,floatfix,amsmath,superscriptaddress,amssymb,bibnotes]{revtex4-1}

\usepackage{graphicx}
\usepackage{bm}
\usepackage{epstopdf}
\usepackage{siunitx}
\usepackage{braket}
\usepackage{pbox}
\usepackage[colorlinks, linkcolor = blue, citecolor = blue, filecolor = black, urlcolor = blue]{hyperref}
\usepackage{xspace} 
\usepackage{units}
\usepackage{amsmath,amssymb}

\DeclareSIUnit\wn{\cm\tothe{-3}}
\newcommand{\rs}{\rm \scriptscriptstyle}

\begin{document}

\title{Electromagnetically induced transparency of ultralong-range Rydberg molecules}

\author{Ivan Mirgorodskiy}
\email{i.mirgorodskiy@physik.uni-stuttgart.de}
\affiliation{5. Phys. Institut and Center for Integrated Quantum Science and Technology, Universit\"at Stuttgart, Germany}
\author{Florian Christaller} 
\author{Christoph Braun} 
\author{Asaf Paris-Mandoki}
\author{Christoph Tresp}
\author{Sebastian Hofferberth}
\email{hofferberth@sdu.dk}
\affiliation{5. Phys. Institut and Center for Integrated Quantum Science and Technology, Universit\"at Stuttgart, Germany}
\affiliation{Department of Physics, Chemistry and Pharmacy, University of Southern Denmark, Odense, Denmark}
\date{\today}

\begin{abstract}
We study the impact of Rydberg molecule formation on the storage and retrieval of Rydberg polaritons in an ultracold atomic medium. We observe coherent revivals appearing in the retrieval efficiency of stored photons that originate from simultaneous excitation of Rydberg atoms and Rydberg molecules in the system with subsequent interference between the possible storage paths. We show that over a large range of principal quantum numbers the observed results can be described by a two-state  model including only the atomic Rydberg state and the Rydberg dimer molecule state. At higher principal quantum numbers the influence of polyatomic molecules becomes relevant and the dynamics of the system undergoes a transition from coherent evolution of a few-state system to an effective dephasing into a continuum of molecular states. 
\end{abstract}

\maketitle

Mapping the long-range interaction between Rydberg atoms \cite{Saffman2010} onto slowly traveling polaritons via electromagnetically induced transparency (EIT) \cite{Fleischhauer2005} has emerged as a promising approach to realize effective photon-photon interaction in an optical medium \cite{Kurizki2005,Kurizki2011,Adams2010,Lukin2011,Vuletic2012}. Fast development over the last years in this novel field of Rydberg quantum optics \cite{Hofferberth2016d} has enabled new tools for quantum information as well as for exploring dynamics of correlated quantum many-body systems \cite{Chang2014}, including efficient single-photon generation \cite{Kuzmich2012b,Adams2013}, creation of entanglement between light and atomic excitations \cite{Kuzmich2013}, realization of attractive forces between single photons \cite{Vuletic2013b}, demonstration of single-photon all-optical switches \cite{Duerr2014} and transistors \cite{Hofferberth2014,Rempe2014b,Hofferberth2016}, single-photon absorbers \cite{Hofferberth2016e} and interaction induced photon phase shifts \cite{Grangier2012,Duerr2016,Vuletic2017}. Future prospects include the crystallization of photons \cite{Pohl2013, Fleischhauer2013} and the observation of three-body interaction between photons \cite{Buechler2016,Gorshkov2016}.

The critical figure of merit for most of the above work is the optical depth (OD) per blockade volume \cite{Lukin2011}. Improving this quantity requires increasing the atomic density of the medium, but this inevitably brings the system into the regime where formation of ultralong-range Rydberg molecules \cite{Sadeghpour2000,Bendkowsky2009} has to be taken into account to describe the Rydberg-polariton dynamics. The experimental study of these exotic molecules has evolved into a highly active field in itself, with a variety of exciting phenomena having been explored so far, such as states bound by quantum reflection \cite{Bendkowsky2010}, coherent creation and breaking of the molecular bond \cite{Butscher2011}, polyatomic Rydberg molecules \cite{Gaj2014}, trilobite \cite{Li2011,Booth2015} and butterfly \cite{Ott2016b} states, and controlled hybridization of the molecular bond \cite{Gaj2015}. Diatomic Rydberg molecules have been realized for $S$-states in Rb \cite{Bendkowsky2009}, Cs \cite{Tallant2012} and Sr \cite{DeSalvo2015}, for $D$-states in Rb \cite{Anderson2014,Krupp2014} and for $P$-states in Rb \cite{Bellos2013} and Cs \cite{Sassmannhausen2015}. Furthermore, Rb$_\text{2}$ Rydberg molecules have been used as a  probe of the quantum phase transition from the superfluid to the Mott-insulator phase \cite{Manthey2015} and the existence of Rydberg molecules bound by mixed singlet-triplet scattering \cite{Anderson2014b} has been proven experimentally for Cs \cite{Sassmannhausen2015} and Rb \cite{Hofferberth2016b,Ott2016}.

In the context of Rydberg quantum optics, formation of Rydberg molecules has been suggested as a limiting factor on the coherence of slow and stored Rydberg polaritons \cite{Gaj2014}. Experimentally, D\"{u}rr \textit{et al.} have observed a dephasing rate of single photons stored in a Rydberg state with large principal quantum number ($n=100$) increasing linearly with atomic density, restricting the overall performance of an all-optical switch \cite{Duerr2014}.

In the present work, we study the storage of Rydberg polaritons systematically over a large range of principal quantum numbers and atomic densities to verify the connection between polariton dephasing and molecule formation. Specifically, we are able to match the binding energy of Rydberg dimers to the period in an observed revival of the retrieval efficiency in an intermediate range of principal quantum numbers $n\approx 50...70$. Our results are well-described by a simple two-species model, including Rydberg atoms and Rydberg dimers, from which we extract the coherence time of photons stored in Rydberg dimers. At higher principal quantum numbers, we observe the transition from coherent dynamics  to an effective dephasing when the number of involved molecular states grows \cite{Gaj2014,Duerr2014}. Experiments and analysis similar to ours have recently been performed by Baur \textit{et al.} \cite{Baur2015}.

\begin{figure}[t]
\includegraphics[width=8.6cm]{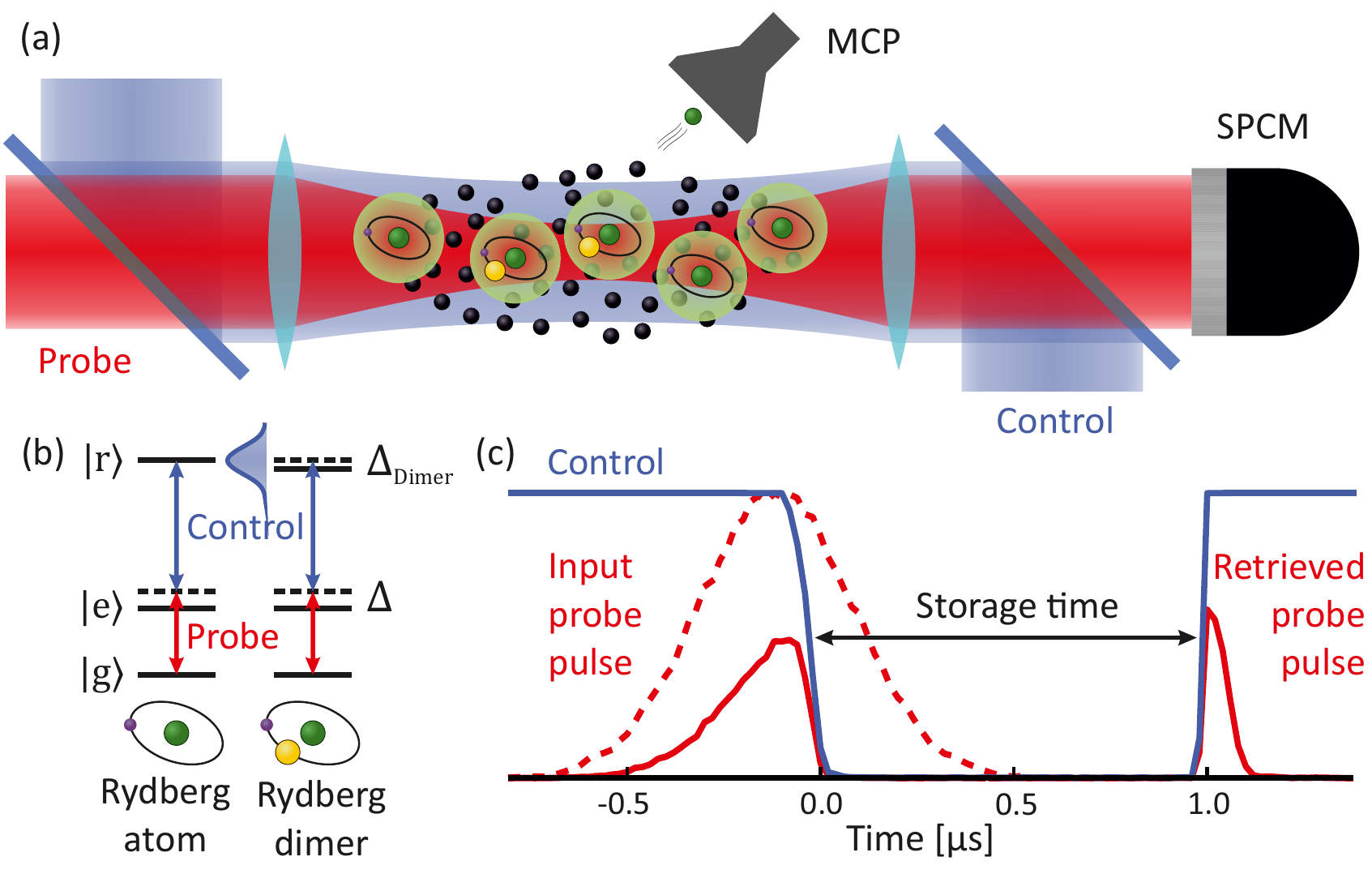}
\caption{\label{fig1:experiment}(a) Scheme of the experiment: a weak \SI{780}{\nano\meter} probe field and a strong \SI{480}{\nano\meter} control field are focused onto a cloud of $^{87}$Rb, coupling the ground state $\Ket{g}$ to an atomic Rydberg state (green atoms) and the slightly shifted Rydberg molecule state (green - yellow atom pairs). (b) Effective two-species level structure for the Rydberg state and the Rydberg dimer with binding energy of $-h \Delta_{\rs{Dimer}}$. If $\Delta_{\rs{Dimer}}$ is smaller than the EIT linewidth both states are coupled simultaneously to the probe photons. (c) Pulse sequence for storage and retrieval of Rydberg polaritons showing control light (blue curve), input probe pulse (red dashed curve), and initially transmitted (solid red pulse on the left) and retrieved  (solid red pulse on the right) probe photons.} 
\end{figure}

A schematic of our experiment is shown in Fig.~\ref{fig1:experiment}. We prepare $9 \times 10^{4}$ $^{87}$Rb atoms, pumped into the ground state $\Ket{g}=\Ket{5S_{1/2},F = 2, m_F = 2}$, trapped in a crossed optical dipole trap. The $1/\text{e}$ radial and axial radii of the cigar shaped cloud at the temperature of \SI{4}{\micro\kelvin} are $\sigma_{\rs \textit{R}} = \SI{13}{\micro\meter}$ and $\sigma_{\rs \textit{L}} = \SI{42}{\micro\meter}$. A weak \SI{780}{\nano\meter} probe field, which couples the ground state $\Ket{g}$ to the intermediate state $\Ket{e}=\Ket{5P_{3/2}, F = 3, m_F = 3 }$, is focused onto the center of the cloud ($w_{\rs 0,{probe}} = \SI{6.4}{\micro\meter}$). We measure $\text{OD} = 24$ on the  $\Ket{g} \rightarrow\Ket{e}$ transition over the full cloud length. For coupling the probe photons to a Rydberg state $\Ket{r}=\Ket{nS_{1/2},m_J=1/2}$ we add a strong \SI{480}{\nano\meter} counterpropagating control field ($w_{\rs 0,{control}} = \SI{14}{\micro\meter}$). To measure Rydberg excitation spectra we detune both probe and control fields by $\Delta =  2\pi \times \SI{100}{\mega\hertz}$ from the intermediate state and then scan the two-photon detuning over the $\Ket{g}$ to $\Ket{r}$ two-photon transition. After the excitation we field-ionize Rydberg atoms and collect ions on a microchannel plate detector (MCP). To perform storage and retrieval experiments, we employ the sequence shown in Fig.~\ref{fig1:experiment}(c). First, we send a Gaussian probe pulse containing on average $n=0.24$ photons into the cloud under resonant conditions ($\Delta = 0$) and ramp the intensity of the control field down to zero on a timescale of \SI{120}{\nano\second} while the probe pulse propagates through the cloud. As a result, a part of the probe pulse is stored as a stopped Rydberg polariton. To read out the Rydberg polariton from the cloud, the control field is turned back on after variable storage time $t_{\rs S}$ and the output optical pulse is collected on a single-photon counting module (SPCM).

\begin{figure}[t]
\includegraphics[width=8.6cm]{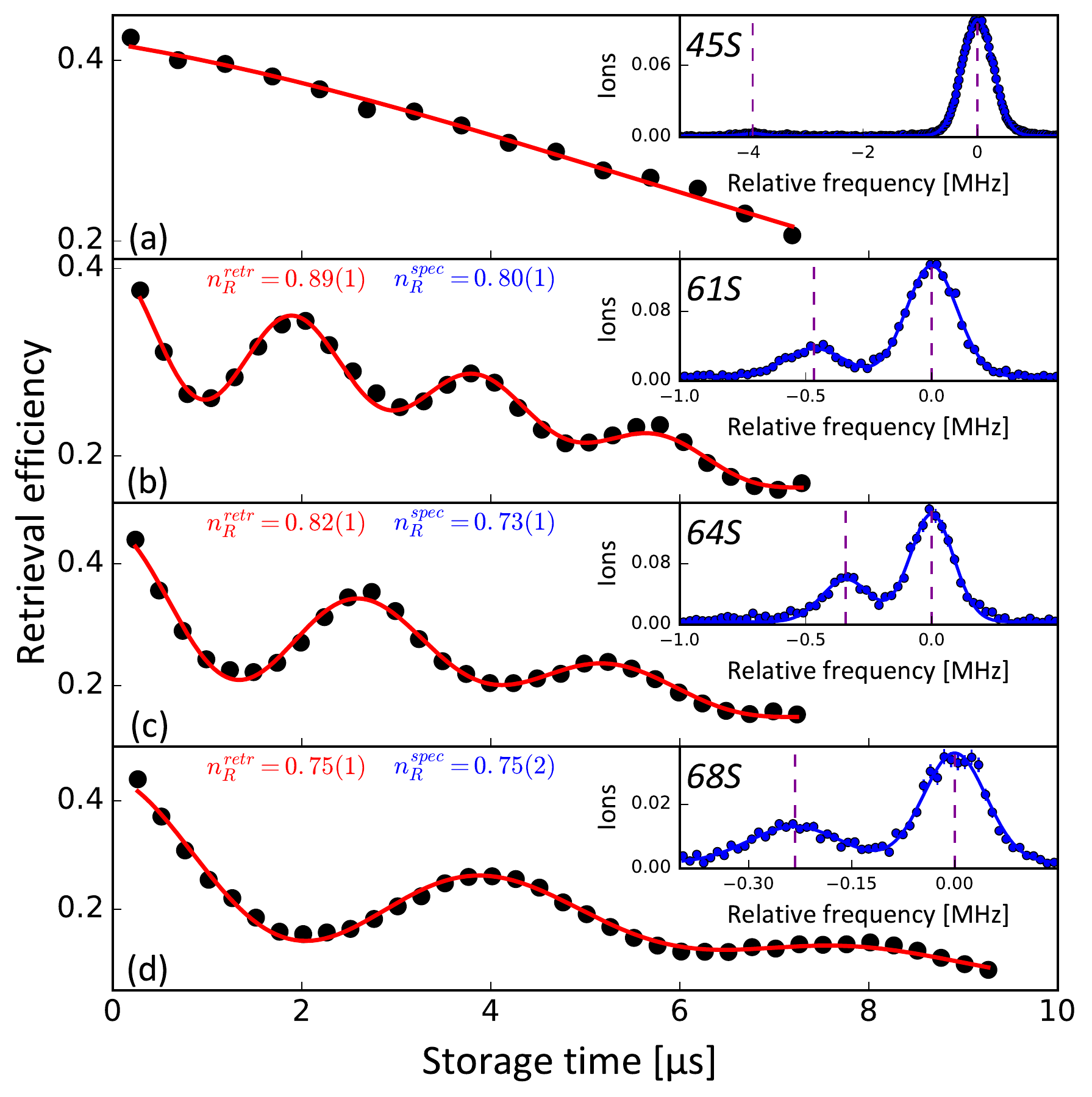}
\caption{\label{fig2:retrieval}Retrieval efficiency of stored photons versus storage time measured for different quantum numbers: (a) $45S$, (b) $61S$, (c) $64S$, (d) $68S$ (black dots). Corresponding Rydberg excitation spectra are shown as insets (blue dots). We fit the excitation spectra with a sum of two Gaussian lineshapes (blue solid line) and independently the retrieval curves with the model discussed in the text (red solid line). From both fits, we obtain the ratio of Rydberg atoms and dimers in the cloud, $n_{\rs \textit{R}}^{\rs \textit{retr}}$ from the fit of retrieval curves and $n_{\rs \textit{R}}^{\rs \textit{spec}}$ from the fit of molecular spectra. The errorbars for most data points are smaller than the shown markers.}
\end{figure}

We start our studies by investigating photon storage and Rydberg molecule formation over a range of principal quantum numbers between $n=45...68$. Fig.~\ref{fig2:retrieval} shows the observed retrieval efficiency plotted versus storage time $t_{\rs S}$ for four different Rydberg states. We observe oscillations in the retrieval efficiency for $61S$, $64S$ and $68S$  with frequencies of $\SI[parse-numbers = false]{511(3)}{\kilo\hertz}$, $\SI[parse-numbers = false]{369(2)}{\kilo\hertz}$ and $\SI[parse-numbers = false]{241(2)}{\kilo\hertz}$. To relate these oscillations to Rydberg dimers, we show the excitation spectra in the corresponding insets in Fig.~\ref{fig2:retrieval}. Each spectrum consists of an atomic peak corresponding to excitation of Rydberg atoms, situated at the origin, and a clearly distinguishable dimer line red-detuned from the atomic line. The underlying mechanism for Rydberg molecule formation is the scattering of the slow Rydberg electron from a ground state atom \cite{Greene2000}. In our case, the observed Rydberg dimers are well-described by pure s-wave scattering \cite{Bendkowsky2009}, resulting in a molecular binding potential
\begin{equation}\label{eq:Potential}
V(\textbf{R}) = \frac{2\pi\hbar^{2}a}{m_{\rs e}}\lvert\Psi(\textbf{R})\rvert^{2},
\end{equation}
where $\lvert\Psi(\textbf{R})\rvert^{2}$ is the local electron density and $\textit{a} \approx -15.7 a_{\rs 0}$ is the s-wave scattering length for the electron - Rb atom collision \cite{Gaj2014}. The vibrational ground state wavefunction of the dimer is mainly localized in the outermost lobe of the Rydberg electron wavefunction, resulting in a characteristic scaling of the dimer binding energy with principal quantum number $n$ \cite{Bendkowsky2009}. The Rydberg electron can also capture multiple ground state atoms, forming polyatomic molecules with binding energy of a molecule with $N$ ground state atoms $(N-1)$ times larger than the binding energy of the dimer \cite{Gaj2014}. In our experiment the mean atomic density over the full cloud is $\SI[parse-numbers = false]{1.4\times10^{12}}{\wn}$, which is sufficient to form a significant fraction of Rydberg dimers, but only a negligible fraction of polyatomic Rydberg molecules for $n \leq 70$. We extract the Rydberg dimer binding energy $E_{\rs b}=-h\cdot\Delta_{\rs{Dimer}}$ (in the following we will refer to $\Delta_{\rs{Dimer}}$ as a binding energy in frequency units) by fitting a sum of two Gaussian lineshapes to the spectra, finding $\SI[parse-numbers = false]{470(35)}{\kilo\hertz}$, $\SI[parse-numbers = false]{341(35)}{\kilo\hertz}$ and $\SI[parse-numbers = false]{240(15)}{\kilo\hertz}$ for $61S$, $64S$ and $68S$ correspondingly, which are in good agreement with the measured retrieval oscillation frequencies. For $n=45$, the dimer fraction is very small while the dimer binding energy $\Delta_{\rs{Dimer}} = \SI{3.93(4)}{\mega\hertz}$ is larger than the Rydberg EIT linewidth, which we set to $\nu_{\rs{EIT}} = 2 \pi \times \SI{2.5}{\mega\hertz}$ FWHM for all datasets. As expected, in this case no oscillations in retrieval efficiency are observed (Fig.~\ref{fig2:retrieval}(a)).

To quantitatively reproduce the measured retrieval curves we employ a simple two-species model for the stored Rydberg polaritons. We consider our atomic cloud to consist of $N_{\rs \textit{R}}$ Rydberg atoms, positioned such that they cannot form a Rydberg molecule, and $N_{\rs \textit{D}}$ pairs of atoms, positioned such that they can form a Rydberg dimer. For Rydberg states where the dimer binding energy is smaller than the EIT linewidth $\nu_{\rs{EIT}}$, each photon is stored in a superposition of collective Rydberg atom and dimer excitations. The wave function of the stored polariton then evolves in time as
\begin{equation}\label{eq:wavefunction}
\ket{\Psi(t)} = \frac{1}{\sqrt{N_{\rs{exc}}}}\left(\sum\limits_{j=1}^{N_{\rs \textit{R}}}\Ket{R_{\rs \textit{j}}} +\sum\limits_{j=1}^{N_{\rs \textit{D}}}e^{-i2\pi\Delta_{\rs{Dimer}}t}\Ket{D_{\rs \textit{j}}}\right).
\end{equation}
Here, $N_{\rs{exc}} = N_{\rs \textit{R}}+N_{\rs \textit{D}}$, $\Ket{R_{\rs \textit{j}}} = \Ket{g_{\rs 1},..,r_{\rs \textit{j}},..,g_{\rs \textit{N}}}$ is a collective state with atom \textit{j} in Rydberg state $\Ket{r}$ and all others in ground state $\Ket{g}$, $\ket{D_{\rs \textit{j}}} = 1/\sqrt{2}\left(\ket{g_{\rs 1},..,(g_{\rs \textit{j}},r_{\rs \textit{j}}),..,g_{\rs \textit{N}}}+\ket{g_{\rs 1},..,(r_{\rs \textit{j}},g_{\rs \textit{j}}),..,g_{\rs \textit{N}}}\right)$ is the symmetrized dimer state with atom pair \textit{j} forming a Rydberg molecule. 
After a storage time $t_{\rs S}$ the retrieval efficiency is then proportional to the overlap of this wave function with the original one at $t=0$ 
\begin{equation}\label{eq:simpleeff}
\eta(t_{\rs S}) ={} \lvert\braket{\Psi(0)|\Psi(t_{\rs S})}\rvert^{2}.
\end{equation}
In addition to the coherent evolution, we take into account several dephasing mechanisms. First, the thermal atomic motion introduces an overall Gaussian decay of the retrieval efficiency with lifetime $\tau_{\rs \textit{T}}=(\Delta kv_{\rs S})^{-1}$, where $\Delta k$ is the sum of the wavevectors of the EIT light fields and $v_{\rs S}= \sqrt{kT/m}$ is the thermal speed of atoms. Additionally, we take into account spontaneous decay of both the Rydberg atoms and Rydberg dimers with lifetimes $\tau_{\rs \textit{R}}$ and $\tau_{\rs \textit{D}}$ respectively. Including these effects in eq.~\ref{eq:simpleeff}, we write the time dependence of the retrieval efficiency of a stored polariton as
\begin{equation}\label{eq:efficiency}
\eta(t) ={} \eta_{\rs 0}e^{{-t^{2}}/{\tau_{\rs \textit{T}}^{2}}} \lvert n_{\rs \textit{R}}e^{-{t}/{\tau^{}_{\rs \textit{R}}}} + n_{\rs \textit{D}}e^{-{t}/{\tau^{}_{\rs \textit{D}}}}e^{-i2\pi\Delta_{\rs{Dimer}}t}\rvert^{2},
\end{equation}
where $\eta_{\rs 0}$ is the initial retrieval efficiency which is determined experimentally; $n_{\rs \textit{R}}$ and $n_{\rs \textit{D}}$ are the fractions of Rydberg atoms and Rydberg dimers, normalized such that $n_{\rs \textit{R}}+n_{\rs \textit{D}}=1$. The oscillations in the retrieval efficiency emerge from eq.~\ref{eq:efficiency} because of the interference between the polariton component stored in the atomic fraction with the one stored in the dimer fraction. We note that because we send on average $n=0.24$ photons per experimental realization, we rarely store more than one photon and thus can neglect any interaction between stored polaritons \cite{Kuzmich2012,Kuzmich2012b,Adams2017}.

We use eq.~\ref{eq:efficiency} to fit the experimental data in Fig.~\ref{fig2:retrieval}. The Rydberg atom lifetime $\tau_{\rs \textit{R}}$ can be calculated and therefore we fix it during the fitting to \SI{49}{\micro\second} ($45S$), \SI{104}{\micro\second} ($61S$), \SI{118}{\micro\second} ($64S$) and \SI{136}{\micro\second} ($68S$). First, for the $45S$ state (Fig.~\ref{fig2:retrieval}(a)), we set $n_{\rs \textit{R}}=1$, which leaves only the thermal dephasing as a free parameter for which we obtain $\tau_{\rs \textit{T}} = \SI[parse-numbers = false]{11.8(3)}{\micro\second}$. This value is consistent with the dephasing time inferred from the cloud temperature $T={}$\SI{4}{\micro\kelvin} measured by absorption imaging, which suggests $\tau_{\rs \textit{T}}$ of \SI{10}{\micro\second}. Thus the time dependence of the retrieval efficiency, when we do not couple to molecular states, is well described by decay due to random thermal motion of atoms and Rydberg population decay.

Next, for the data shown in Fig.~\ref{fig2:retrieval}(b)-(d), we keep $\tau_{\rs \textit{T}}$, $\tau_{\rs \textit{D}}$, $n_{\rs \textit{R}}$ and $\Delta_{\rs{Dimer}}$ as free parameters. The resulting values for $\Delta_{\rs{Dimer}}$ are in very good agreement with the values we get from the excitation spectra (see above). Lifetimes $\tau_{\rs \textit{T}}$ of thermal atomic motion extracted from the fits $\SI[parse-numbers = false]{10.0(2)}{\micro\second}$ ($61S$), $\SI[parse-numbers = false]{9.6(2)}{\micro\second}$ ($64S$) and $\SI[parse-numbers = false]{9.9(2)}{\micro\second}$ ($68S$), all of which are again in very good agreement with the measured cloud temperature. The parameters defining the contrast of the observed oscillations, are the fraction of Rydberg atoms $n_{\rs \textit{R}}$ and the fraction of Rydberg dimers $n_{\rs \textit{D}}=1-n_{\rs \textit{R}}$. For our set of data, the fit yields a Rydberg fraction of 0.89(1) ($61S$), 0.82(1) ($64S$) and 0.75(1) ($68S$). When we compare these values to the ones obtained from the amplitudes of the Gaussian fits to atomic and molecular lines in the excitation spectra, 0.80(1) ($61S$), 0.73(1) ($64S$), 0.75(2) ($68S$), we observe discrepancies for the $61S$ and $64S$ states, exceeding one standard deviation. It is important to note that the result of the fit is very sensitive to the value of $n_{\rs \textit{R}}$ and it is not possible to compensate an offset from the optimum value of $n_{\rs \textit{R}}$ by adjusting other fit parameters. Instead, the observed disagreement in the parameters is explained by systematic drifts of the total atom number, which over the time required for the two measurements is on the order of 10\% in our setup. For the shown datasets we can confirm these drifts by evaluating control measurements of optical depth taken throughout the full measurement sequence, which confirm the change in atom number between the spectra and retrieval data sets. 
Finally, we extract the Rydberg dimer coherence times $\SI[parse-numbers = false]{4.5(6)}{\micro\second}$ ($61S$), $\SI[parse-numbers = false]{4.3(4)}{\micro\second}$ ($64S$) and $\SI[parse-numbers = false]{4.7(3)}{\micro\second}$ ($68S$). These values are comparable to the Rydberg dimer lifetimes measured by other methods in previous experiments \cite{Butscher2010,Butscher2011,Ott2015}. While our method does not differentiate between decay time $T_{\rs 1}$ and dephasing time $T_{\rs 2}$ \cite{Butscher2010}, our results suggest that coherent photon storage in the Rydberg dimer state is possible over time scales required for conditional photon-photon interaction \cite{Duerr2014,Hofferberth2016,Duerr2016}.

\begin{figure}[t]
\includegraphics[width=8.6cm]{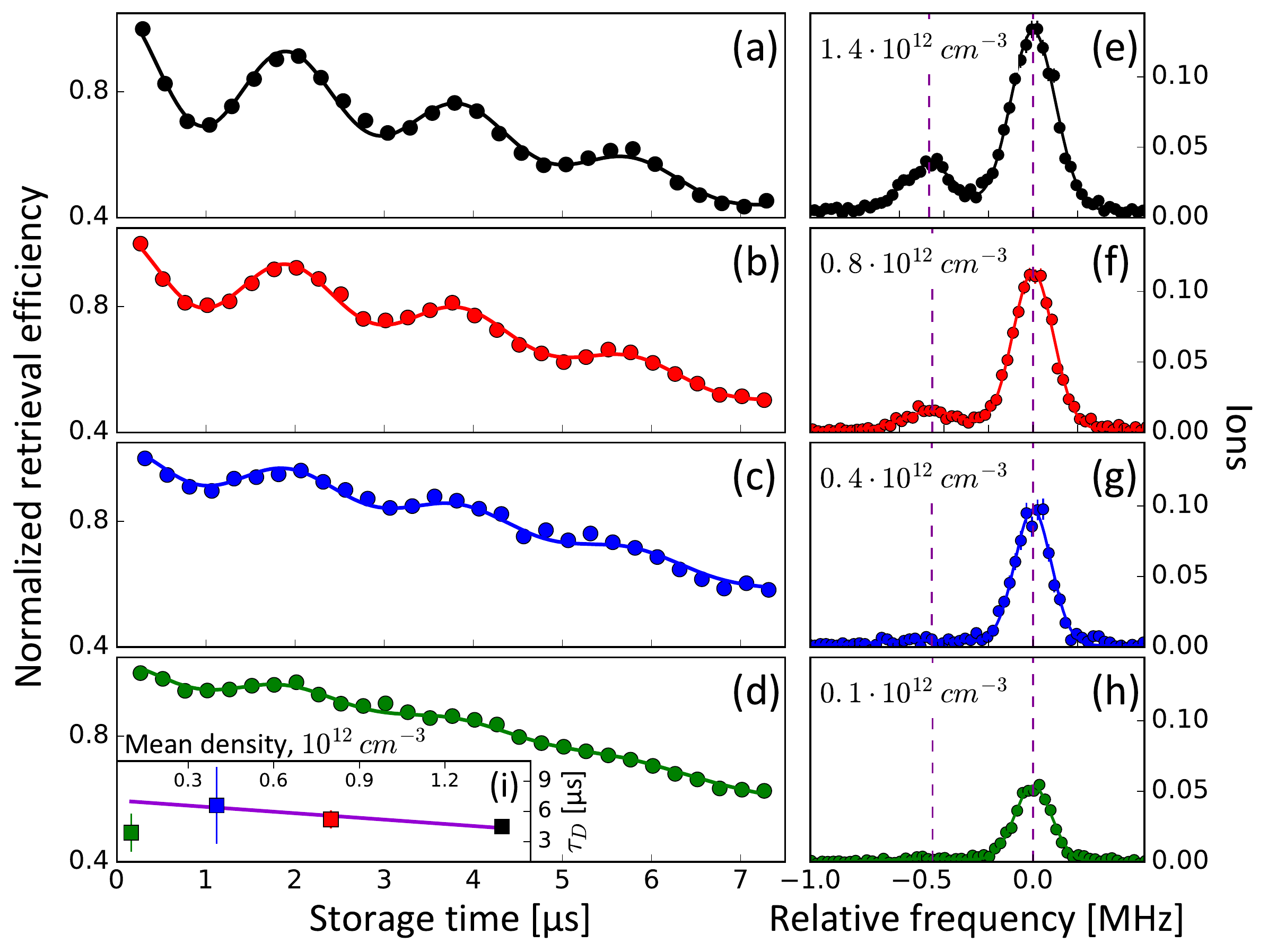}
\caption{\label{fig3:density} (a)-(d) Retrieval efficiency of stored photons versus storage time measured for different densities (high to low) for the $61S$ state (dots). Solid lines are fits to the data using eq.~\ref{eq:efficiency}. (e)-(h) Corresponding excitation spectra from which the indicated mean atomic density for each data set is extracted. The contrast of the retrieval efficiency oscillation decreases as expected when the amplitude of the dimer line becomes smaller. (i) Rydberg dimer coherence times for the different atomic densities extracted from the fits to the curves in (a)-(d).}
\end{figure}

We next turn to investigating the time dependence of the retrieval efficiency for different atomic densities, which allows to alter the Rydberg dimer fraction $n_{\rs \textit{D}}$. Fig.~\ref{fig3:density} shows the measured retrieval efficiency curves at four atomic densities for the $61S$ state as well as the corresponding Rydberg excitation spectra. We extract the mean atomic density for each dataset from the position of the center of gravity of the excitation spectrum \cite{Gaj2014}, obtaining $\SI[parse-numbers = false]{1.4\times10^{12}}{\wn}$, $\SI[parse-numbers = false]{0.8\times10^{12}}{\wn}$, $\SI[parse-numbers = false]{0.4\times10^{12}}{\wn}$ and $\SI[parse-numbers = false]{0.1\times10^{12}}{\wn}$ (Fig.~\ref{fig3:density}(e)-(h)). Correspondingly, the Rydberg dimer fraction $n_{\rs \textit{D}}$ shrinks with the density down to $n_{\rs \textit{D}}=0.02$ for the lowest density. Therefore, as expected from eq.~\ref{eq:efficiency}, we observe a decreasing contrast in the retrieval efficiency oscillations when lowering the density (Fig.~\ref{fig3:density}(a)-(d)). More quantitatively, by fitting the data with the model, we extract the Rydberg atom fractions $n_{\rs \textit{R}}= 0.89(1), 0.94(1), 0.97(1), 0.98(1)$ for the four datasets, which are again in agreement with the ratios extracted from the excitation spectra within our experiment stability. In Fig.~\ref{fig3:density}(i) we show the Rydberg dimer coherence times extracted from the fits for the different atomic densities. Omitting the result for the lowest density, where the dimer fraction is very small and the oscillations are barely visible, the remaining data points suggest a linear decrease of the coherence time as was found previously by Baur \textit{et al.} \cite{Duerr2014}.

\begin{figure}[t]
\includegraphics[width=8.6cm]{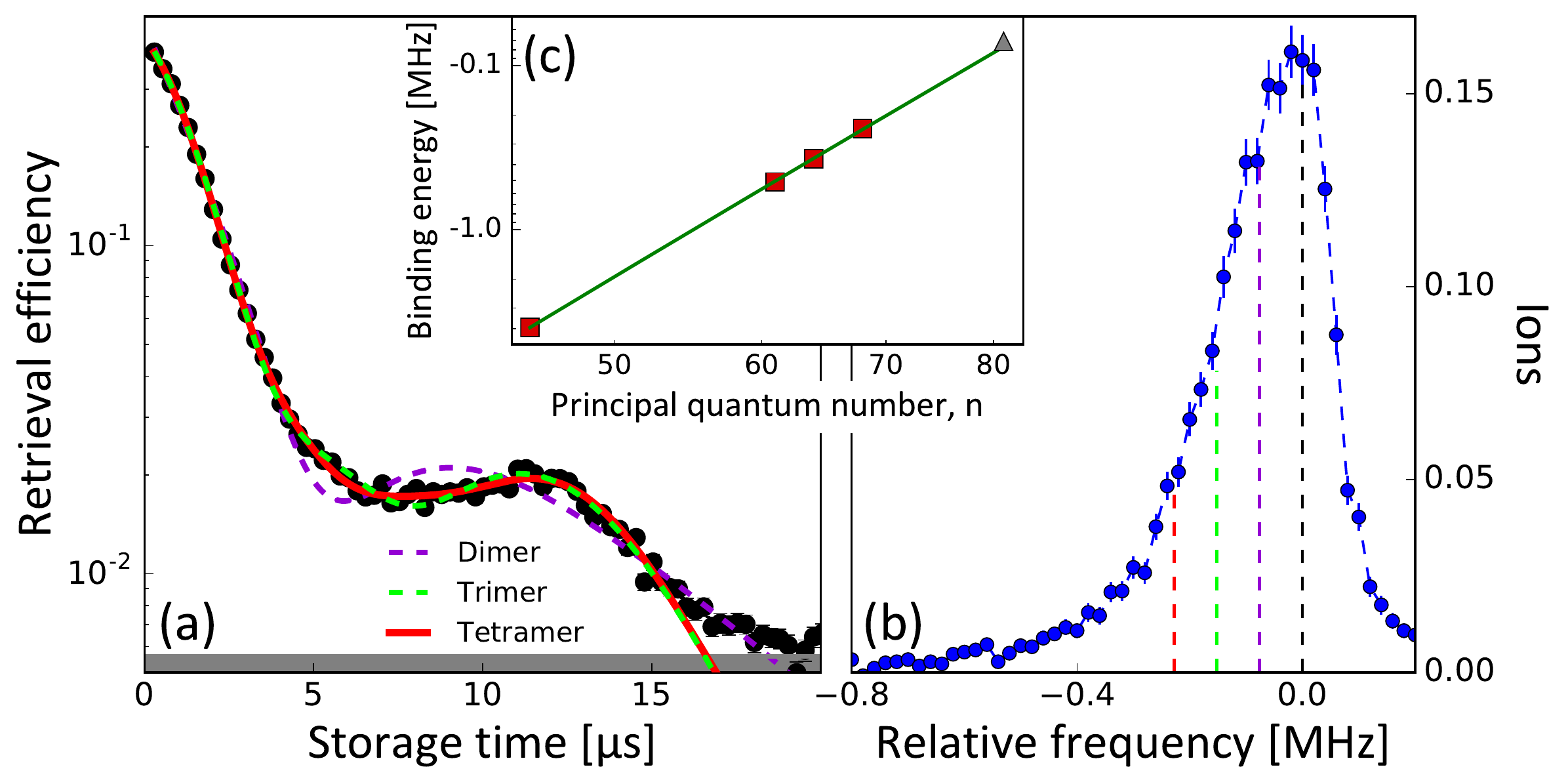}
\caption{\label{fig4:81S}(a) Retrieval efficiency vs storage time for $81S$ (in semilog scale). The lines are fits to the experimental data (black dots) including polyatomic molecule states up to the size indicated in the figure. (b) Corresponding excitation spectrum for $81S$. Dashed lines mark the expected centers of atomic, dimer, trimer and tetramer lines (from the right to the left), based on the dimer binding energy extrapolated from the data at lower $n$. (c) Measured binding energies versus principal quantum number. Red squares are the results for $n=45,61,64,68$ used for the fit (green solid line). Grey triangle is the binding energy extracted from the fit to the retrieval efficiency in (a) including molecule states up to tetramer.}

\end{figure}
Next, we investigate photon storage in a Rydberg state with larger principal quantum number, specifically $81S$, to highlight the transition from coherent dynamics to effective dephasing observed previously at $n=100$ \cite{Duerr2014}. For the $81S$ state the dimer lineshift is already smaller than the dimer linewidth, resulting in an asymmetric Rydberg line in the excitation spectrum with a non-resolvable shoulder on the red-detuned side (Fig.~\ref{fig4:81S}(b)). From the spectrum, we can thus no longer  directly determine the dimer binding energy. This can only be done indirectly by determining the scaling of the binding energy with  effective quantum number $n^{\rs{*}}$ from the data at lower $n$ (Fig.~\ref{fig4:81S}(c)). Fitting a power law $\Delta_{\rs{Dimer}} \sim \left(n^{\rs{*}}\right)^{\rs{\alpha}}$ to the binding energies extracted for $45S$, $61S$, $64S$ and $68S$, we obtain $\alpha = -6.35(4)$, which is in very good agreement with the value $-6.26(12)$ found by Gaj \textit{et al.} \cite{Gaj2014}. Using the binding energy $\Delta_{\rs{Dimer}} = \SI{76}{\kilo\hertz} $ predicted from this scaling, we mark the line centers of the atomic, dimer, trimer and tetramer contributions in the spectrum shown in Fig.~\ref{fig4:81S}(c). From these line positions it becomes clear that for $81S$ and our atomic density $\SI[parse-numbers = false]{1\times10^{12}}{\wn}$, there should already be a sizable contribution from Rydberg trimers and even tetramers in the photon storage process. Consequently, our two-species model fails to adequately describe the observed revival in the retrieval efficiency in this case (dashed violet line in Fig.~\ref{fig4:81S}(a)). We can significantly improve our model by adding terms corresponding to Rydberg trimers (dashed green line) and tetramers (red line) with binding energies multiples of the dimer binding energy to eq.~\ref{eq:efficiency}. The fit with a total of four species (atoms, dimers, trimers, tetramers) reproduces the single revival feature very well, adding further polymer states does not improve the fit. Importantly, from this fit we obtain the dimer binding energy as $\Delta_{\rs{Dimer}} =  \SI{71(1)}{\kilo\hertz}$ (grey triangle in Fig.~\ref{fig4:81S}(c)), in very good agreement with the value predicted by the scaling law. This shows that the photon storage and retrieval is a useful method to measure Rydberg molecule binding energies even if the molecular lines are not resolvable in the excitation spectrum.

In conclusion, we have investigated the coherent oscillations appearing in the retrieval efficiency of stored photons in experiments on storage and retrieval of Rydberg polaritons. We explain these oscillations by simultaneous excitation of Rydberg atoms and Rydberg dimers in the system with subsequent interference between the possible storage paths. Our observations are well reproduced by a simple model including only the Rydberg dimer state over a range of principal quantum numbers $n\approx50...70$. For higher principal quantum numbers, more molecular states become relevant. We show that our model still works well for $81S$ if the trimer and tetramer vibrational ground states are included. At even higher principal quantum number, the density of molecular states increases so much, that the retrieval evolution is well described by a single decay term \cite{Duerr2014}. A more sophisticated model would require including larger polyatomic molecules as well as vibrational excited states of the Rydberg molecules \cite{Demler2016}. From the perspective of Rydberg quantum optics, our observation of coherent light storage in Rb Rydberg dimers with principal quantum numbers close to the two-state F\"{o}rster resonance exploited for efficient all-optical switch and transistor operation \cite{Rempe2014b,Hofferberth2016,Hofferberth2016f} should prove highly relevant for the further improvement of these devices and for the realization of coherent photonic gates based on Rydberg-mediated photon-photon interaction.

\begin{acknowledgments}
We acknowledge funding by the German Research Foundation (Emmy-Noether-grant HO 4787/1-1, GiRyd project HO 4787/1-3, SFB/TRR21 project C12) and the Ministry of Science, Research and the Arts of Baden-W\"{u}rttemberg (RiSC grant 33-7533.-30-10/37/1)
\end{acknowledgments}

%

\end{document}